\documentclass[journal]{IEEEtran}
\usepackage{amsmath,amsfonts}
\usepackage{algorithmic}
\usepackage{algorithm}
\usepackage{array}
\usepackage[caption=false,font=normalsize,labelfont=sf,textfont=sf]{subfig}
\usepackage{textcomp}
\usepackage{stfloats}
\usepackage{url}
\usepackage{verbatim}
\usepackage{graphicx}
\usepackage{cite}
\usepackage{xcolor}
\newif\ifshowrev{}
\showrevfalse{}

\ifshowrev
  \newcommand{\revA}[1]{{\color{red}#1}}%
  \newcommand{\revB}[1]{{\color{blue}#1}}%
  \newcommand{\revBoxB}[1]{{%
      \setlength{\fboxrule}{1pt}%
      \setlength{\fboxsep}{0pt}%
      \fcolorbox{blue}{white}{#1}%
    }}
\else
  \newcommand{\revA}[1]{#1}%
  \newcommand{\revB}[1]{#1}%
  \newcommand{\revBoxB}[1]{#1}%
\fi

\begin{document}

\bstctlcite{IEEEBST:remove_url_prefix} % Load the custom control

\title{Beyond Redundancy: Toward Agile Resilience in Optical Networks to Overcome Unpredictable Disasters}

\author{Toru~Mano, Hideki~Nishizawa, Takeo~Sasai, Soichiroh Usui, Dmitrii~Briantcev, Devika~Dass, Brandt~Bashaw, Eoin~Kenny, Marco~Ruffini, Yoshiaki~Sone, Koichi~Takasugi, and Daniel~Kilper
  %\vspace{-10mm}
}

\maketitle

\begin{abstract}
Resilience in optical networks has traditionally relied on redundancy and pre-planned recovery strategies, both of which assume a certain level of disaster predictability.
However, recent environmental changes such as climate shifts, the evolution of communication services, and rising geopolitical risks have increased the unpredictability of disasters, reducing the effectiveness of conventional resilience approaches.
To address this unpredictability, this paper introduces the concept of agile resilience, which emphasizes dynamic adaptability across multiple operators and layers.
We identify key requirements and challenges, and present enabling technologies for the realization of agile resilience.
Using a field-deployed transmission system, we demonstrate rapid system characterization, optical path provisioning, and database migration within six hours.
These results validate the effectiveness of the proposed enabling technologies and confirm the feasibility of agile resilience.
\end{abstract}


%\vspace{-6mm}
\section{Introduction}
\label{sec:introduction}

\IEEEPARstart{N}{etwork} resilience is a fundamental requirement for ensuring service continuity for society and businesses.
Even when infrastructure is damaged by natural disasters such as earthquakes, tsunamis, or floods, network services are expected to remain available.
Conventional approaches to achieving resilience include (a) enhancing reliability through redundancy, (b) securing critical communications via congestion control and prioritized routing, and (c) enabling rapid restoration of services in accordance with predefined protocols~\cite{agrawal2019network}.
% ~\cite{comprehensive_survey, challenges_opportunities,agrawal2019network}. To save no. of refs.
These approaches typically rely on estimating disaster impact and service demand from demographic and facility locations; the importance of each facility is identified along with resource availability as key inputs for network design and disaster response.
These approaches implicitly assume that traffic demand, facility location, and disaster scenarios can reasonably predictable.
For instance, traffic demand, largely driven by human-to-human communication, has historically been inferred from demographic data, which are relatively stable and predictable.
Network facility locations also tend to remain unchanged over decades of operation, while natural disasters have been predictable to some extent based on historical records.
Under such conditions, these assumptions were regarded as valid.

In recent years, however, the dominant type of service, climatic conditions, and geopolitical risks have undergone significant change, undermining the validity of the conventional assumptions of predictable traffic demand, facility locations, and disaster effects.
Climate change due to global warming has led to unprecedented events, including floods and wildfires that cannot be explained by historical data~\cite{grant2025global}.
In parallel, growing global awareness of potential risks facing infrastructure, including those related to military activities, which can co-occur in multiple locations~\cite{sapotage}, necessitates considering human-induced disruptions in addition to natural disasters.
Such activities take into account redundancy-based disaster countermeasures and attempt to compromise infrastructure by exploiting the weakest points or targeting multiple locations simultaneously.
Therefore, redundancy measures such as backup paths alone are insufficient to ensure resilience.
Another significant change is that machine-to-machine communications have emerged as the dominant form of traffic.
%~\cite{traffic}. % to save reference count
Their demand is determined by the location of data centers hosting vast computational resources rather than by demographic data.
While the prevailing trend once favored the construction of a few massive data centers, floor space and power supply constraints are now driving a shift toward a distributed data center architecture composed of multiple mid-sized data centers.
As services increasingly depend on computing resources, resilience must be extended beyond networking to include computing infrastructures as well.
At the same time, legal frameworks around the world are rapidly evolving to impose stricter data localization requirements and cross-border transfer restrictions, thereby generating regulatory uncertainty, which increase the difficulty of long-term infrastructure planning and investment.
These emerging factors highlight the necessity of rethinking resilience approaches that rely on assumptions of predictability.

To address the increasing unpredictability and scale of disasters, resilience must encompass not only static robustness within a single domain but also dynamic agility across multiple domains.
Implementing comprehensive redundancy to cover all possible disasters is impractical from a cost perspective; therefore, redundancy is typically applied selectively based on disaster predictions.
However, as disaster unpredictability continues to increase, the effectiveness of such pre-disaster countermeasures diminishes, and the need for rapid post-disaster recovery becomes more critical.
In such cases, it is essential to aggregate available resources that remain unaffected by the disaster and dynamically interconnect them to restore services quickly.
Moreover, as the scale of disasters grows, there is no guarantee that a single operator can secure the necessary resources for rapid recovery.
In fact, Japanese telecommunication operators jointly launched the new cooperative framework to share facilities for rapid restoration in December 2024.
% may cut to save words
\revB{%
  This was further demonstrated in March 2025 through a field drill where four mobile operators installed base-station equipment on the cable-laying vessel of a Japanese operator to provide wireless coverage from the sea~\cite{ntt_2025_japan}.
  This exercise served as a concrete proof-of-concept of cross-carrier coordination and asset sharing for emergency network restoration.
  Similarly, in Australia, Telstra has collaborated with the government and other carriers to simulate temporary roaming on its network for users affected by outages~\cite{telstra_2024_temporary}.
}% 
We also note that service continuity requires not only network resources but also computing resources.
Consequently, leveraging resources distributed across multiple operators and layers, rather than relying solely on a single operator or a single layer, can significantly reduce the time required for service restoration.
This paradigm shift requires us to develop agile resilience capabilities, enabling rapid adaptation to unforeseen failures and dynamic resource reallocation across multiple layers and operators.

Although extensive studies have explored resilience techniques such as redundancy in optical networks and replication in distributed systems, these approaches remain limited because they either rely on pre-provisioned resources or operate within a single provider or layer, making them insufficient for achieving full agile resilience against large-scale or unpredictable disasters.
Research on ring protection for dense wavelength division multiplexing (DWDM) transport~\cite{ring_protection} has been conducted for many years and provides high resilience with very short switching times, while multi-layer recovery has also been studied extensively~\cite{multilayer_recovery}.
However, both approaches depend on securing backup paths and equipment in advance, which works for predictable failures but is less effective for unforeseen situations.
Similarly, considerable research on large-scale distributed service design has focused on commercial systems with embedded fault-tolerance mechanisms.
The Google File System~\cite{ghemawat2003gfs}, for example, partitions files into chunks replicated across servers; in cloud environments, region replication establishes redundancy across geographically dispersed sites.
Yet such methods usually assume users can access an alternative region after a large-scale disaster.
% cut to save word count
%Optical Spectrum as a Service (OSaaS) is a model that shares fiber among multiple operators~\cite{osaas}. It has been shown to optimize quality of transmission (QoT) using the channel-proving method; however, it is challenging to optimize line systems that degrade during disasters because the QoT optimization approach of OSaaS is based solely on bit error rate (BER) counts from transceivers.

The contributions of this paper are as follows:
\begin{itemize}
  \item We highlight that recent environmental shifts --- climatic changes, the evolution of communication services, and increasing geopolitical risks --- indicate the need for dynamic agility across multiple domains alongside traditional resilience (Sec.~\ref{sec:introduction}).
  \item Through an example recovery scenario, we identify the requirements and challenges of agile resilience (Sec.~\ref{sec:requirements-and-challenges}), introduce key enabling technologies for its realization (Sec.~\ref{sec:key-enablers}),
  and suggest several directions for future research (Sec.~\ref{sec:future-researches}).
  \item We conduct a proof-of-concept demonstration using a field-deployed transmission system spanning several hundred kilometers, showing the dynamic relocation of network and computing resources, thereby validating both the effectiveness of the key technologies and the feasibility of  agile resilience (Sec.~\ref{sec:field-evaluations}).
\end{itemize}

\section{Agile resilience requirements and challenges}
\label{sec:requirements-and-challenges}

This section introduces an example disaster scenario (Fig.~\ref{fig_crop}) to elucidate the requirements and challenges.

\subsection{Example Disaster Scenario}
\label{sec:example-disaster}

\begin{figure}[!t]
  \centering
  \revBoxB{\includegraphics[width=0.8\columnwidth]{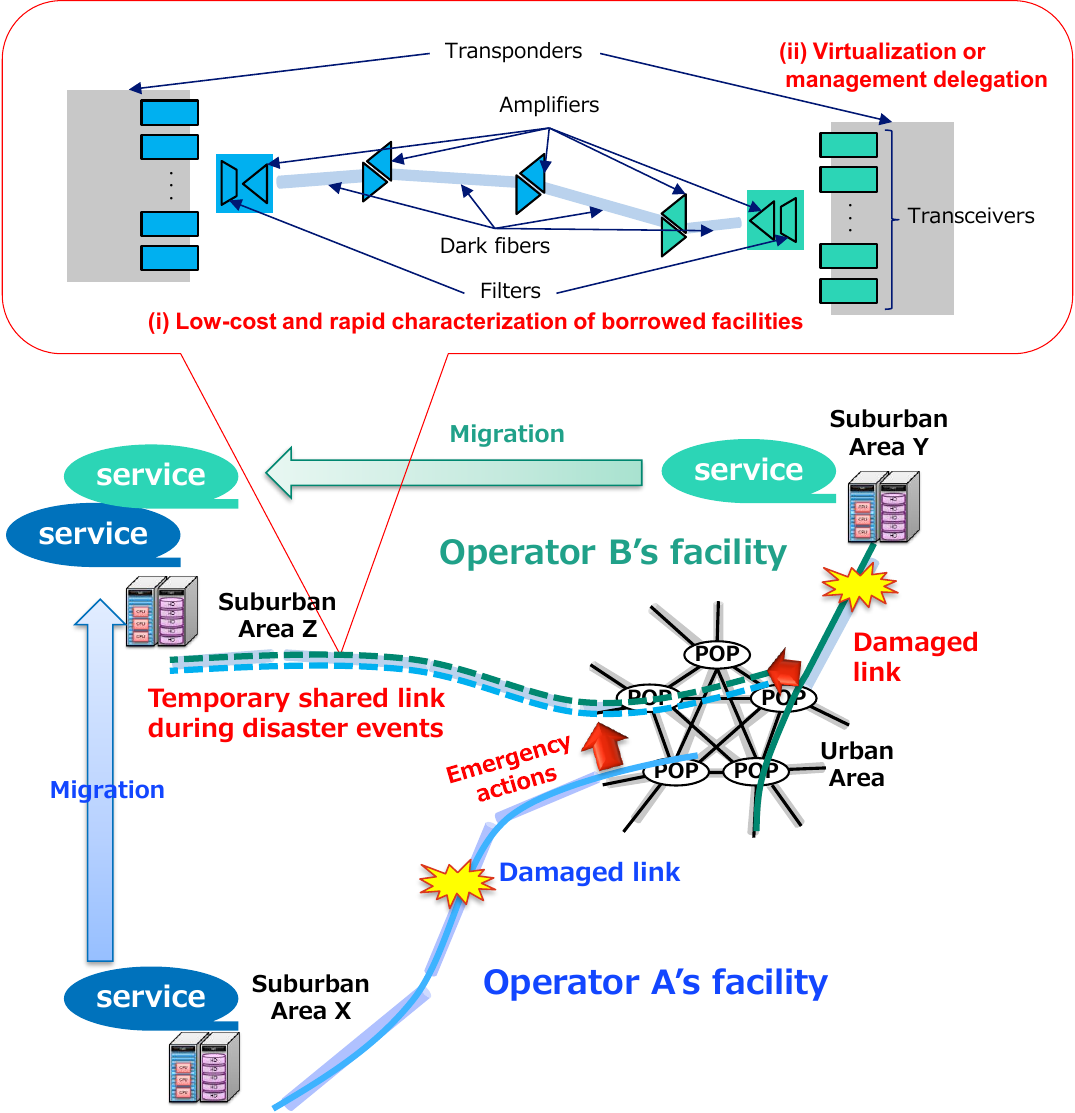}}
  %\vspace{-2mm}
  \caption{Example disaster scenario}
  %\vspace{-6mm}
  \label{fig_crop}
\end{figure}

Operator A operates data centers in suburban area X and an optical network in an urban area (Fig.~\ref{fig_crop}).
They store confidential information (such as financial, medical, and health data) in databases, and use computing resources (e.g., CPUs and GPUs) to provide critical services primarily to urban users.
Operator B also operates an optical network in the same urban area and
\revB{%
  provides critical services via data centers in suburban area Y.
}%
Both operators own optical transmission equipment in the urban area and share several Points of Presence (PoPs).

We consider an unexpected, large-scale natural disaster, such as a massive earthquake or tsunami, that causes widespread damage across
\revB{%
  area X and area Y, resulting in power outages to data centers and optical transmission equipment within 72 to 96 hours.
  While pre-disaster preparedness measures prevent direct damage to the data centers and optical transmission equipment of both operators, the regional power grids or generation facilities suffer severe damage and cease operation.
  To mitigate such power outages, both operators have deployed emergency backup power generators.
}%
The backup generators supply power to the facilities immediately after the power outage.
\revB{%
  However, following the Tier 3 and Tier 4 requirements of ANSI TIA-942, the stored fuel will run out in 72 to 96 hours.
}%

As an emergency measure,
\revB{%
  Operators A and B temporarily borrow data center Z resources in an area suffering relatively minor disaster impact, and cooperatively establish optical paths to the data center and attempt to migrate critical services from the affected areas while backup power is available.
  First, Operators A and B cooperatively establish a physical route between data center Z and the urban area by connecting their dark fibers.
  Next, they deploy spare optical transmission equipment (such as optical amplifiers and reconfigurable optical add \& drop multiplexers (ROADMs)) --- secured from their own central offices or maintenance depots, or rapidly provided from their procurement vendors --- to the communication buildings along the fiber route to construct a temporary line system.
  Using this newly built line system, Operators A establishes optical paths between data center Z and area X.
  While backup power remains available, Operator A migrates essential computing resources, including databases, from area X to data center Z, ensuring service continuity from data center Z.
  Similarly, Operator A establishes additional optical paths between data center Z and area Y to migrate and maintain Operator B's service as well.
}%
Because large-scale disasters simultaneously reduce available infrastructure capacity and increase communication demand for response and recovery operations, the rapid provisioning of high-capacity optical paths
\revB{%
  on newly built temporary infrastructure
}%
is preferred to relying solely on existing IP routing links.
By rapidly establishing new optical paths over
\revB{%
  the ad-hoc infrastructure by leveraging spare assets
}%
and migrating computing resources, Operator A and B can enhance service continuity under catastrophic conditions, thereby improving their resilience.

\subsection{Requirements and Challenges}
\label{sec:requirements-challenges}

To achieve agile resilience for rapid service migration, the following are requirements:

\begin{itemize}
  \item \emph{(1) Rapid establishment of optical paths across \revB{cooperatively built ad-hoc infrastructure}}:
        Operators must be able to \revB{quickly provision optical paths on a newly built line system that leverages spare assets, such as dark fibers and spare equipment, across multiple administrative domains}.
  \item \emph{(2) Temporary sharing of partially idle network and computing resources}:
        Even during a disaster, resources such as fiber spectrum (e.g., unused portions of the C-band), CPU, storage, and switching capacity may remain partially available.
        Temporarily lending such resources to other operators is essential for service relocation and restoration.
\end{itemize}

At least two technical challenges exist to meet the above two requirements (Fig.~\ref{fig_crop}):

\subsubsection*{(i) Low-cost, rapid characterization of borrowed optical transmission facilities}
Accurate estimation of Quality of Transmission (QoT) is required to select appropriate modulation formats and launch powers for new wavelength paths.
QoT depends on the characteristics of system components such as fibers, amplifiers, filters, and transceivers.
Under normal conditions, these parameters (e.g., attenuation coefficient, chromatic dispersion, and nonlinear coefficient) are known from prior measurements.
\revB{%
In the agile resilience scenarios considered here, a new temporary line system will be constructed using dark fibers and spare maintenance equipment; the characteristics of such unused resources are not monitored continuously, as they are typically measured only prior to use.
Even if previous measurement records exist, physical characteristics may have changed due to aging or the impact of the disaster itself.
}%
Rapid path establishment therefore requires quick and accurate extraction of these parameters.
Because measurement instruments and field personnel may be limited during the recovery phase, methods that minimize dedicated instruments and on-site intervention are highly desirable.

\revA{%
  To address this challenge, we introduce Digital Longitudinal Monitoring (DLM) and Optical Line System (OLS) calibration.
  These technologies enable the rapid extraction of fiber and amplifier parameters while using only the transponders and built-in power meters at the path endpoints, eliminating the need for dispatching field personnel or employing dedicated measurement instruments.
}%

\subsubsection*{(ii) Transponder virtualization and/or delegated management of line ports}
Delivering end-to-end services requires both networking and computing resources.
Optical networking resources can be partially leased as spectrum segments, and computing/IP–Ethernet resources can be partially leased via virtualization mechanisms (e.g., VLAN, VXLAN).
The remaining bottleneck is the transponder, which bridges optical resources to computing and packet networks.
Unlike spectrum or compute resources, a transponder cannot be leased in fractions such as by wavelength, and no widely adopted virtualization mechanism currently allows delegation of a portion of a transponder’s capacity.
As a result, it is difficult to temporarily lend unused transponder ports for disaster recovery without transferring full control of the device.

By combining transponder virtualization with existing virtualization technologies, temporary resource lending becomes feasible.
Furthermore, integrating rapid transmission system characterization with fast path design
%~\cite{curri2022gnpy} to save refs.
enables swift optical path provisioning~\cite{nishizawa2024fast}, thereby offering a promising path toward agile resilience.

\section{Key enablers for agile resilience}
\label{sec:key-enablers}

This section reviews state-of-the-art technologies related to agile resilience.
First, we introduce fast and low-cost methods for the characterization of borrowed transmission facilities, including fibers, amplifiers, and transceivers.
Next, we outline peripheral technologies associated with transponder virtualization.

\subsection{Facilities characterization method}

This subsection introduces DLM for fiber characterization, OLS calibration for amplifier characterization, and transceiver noise estimation for transceiver characterization.

\subsubsection{Digital Longitudinal Monitoring}

DLM is a link characterization method that, from just the transponders, estimates end-to-end, fiber-longitudinally distributed optical power and fiber parameters such as attenuation coefficient and chromatic dispersion~\cite{sasai2024linear}.  % to save number of refs.
This approach involves transmitting a signal over the target line system, capturing it at the receiver-side digital signal processor, and extracting the signal power information through analysis of the self-phase modulation imprinted on the received signal.
The longitudinal signal power provides parameters essential for transmission design, such as fiber launch power, the number and location of amplifiers, and the presence of lumped losses.

Compared with conventional optical time-domain reflectometry (OTDR)-based fiber characterization, DLM significantly reduces both the cost and operational effort. Because the reflected signals of OTDR cannot pass through amplifiers due to the presence of isolators, we must dispatch technicians to each amplifier site to measure fiber characteristics across the entire line system.
In contrast, DLM needs only the transponders at the optical path endpoints, eliminating the need for dedicated equipment and on-site technicians, thus making the measurement process more cost-efficient and time-effective.
%In this work, we used the algorithm described in\cite{sasai2024linear} for DLM.

\subsubsection{OLS calibration}

OLS calibration is a method for estimating amplifier characteristics, such as noise figure and frequency-dependent gain, by analyzing amplifier input/output power and the transmitted/received WDM spectra at the optical path endpoints~\cite{borraccini2025optical}.
After characterization, OLS calibration also optimizes amplifier configurations to maximizes QoT based on obtained parameters.
OLS calibration transmits a full C-band WDM comb, changes the gain and tilt settings of amplifiers in the line system, and records amplifier input/output powers and the changes in the WDM spectrum.
By taking into account noise accumulation at amplifiers, spectral variations induced by amplifier gain, and waveform distortions caused by fiber nonlinear effects, we can infer amplifier characteristics from the measured spectral data.
We measure amplifier input/output powers using built-in power meters, and WDM spectrum using the optical spectrum analyzer (OSA) installed at each path endpoint.
%
% cut to save word counts
% The full C-band WDM comb can be generated using an Amplified Spontaneous Emission (ASE) noise source together with the wavelength selective switch (WSS) in the reconfigurable optical add \& drop multiplexer (ROADM), or, in some cases, by a ROADM equipped with a dummy WDM waveform generator.
%
% Such a function is often included to stabilize nonlinear optical effects and amplifier behaviors against changes in the number of WDM signals.
%
Compared with a conventional method in which technicians carry Amplified Spontaneous Emission (ASE) sources and OSAs to each site for amplifier characterization, OLS calibration requires equipment only at the endpoints of the optical path, thereby reducing both cost and operational effort.

\subsubsection{Transceiver noise estimation}

This method estimates transceiver noise using only a single variable optical attenuator (VOA)~\cite{mano2024measuring}.
Transceivers also affect signal quality, and previous studies have shown that their impact can be modeled as additive Gaussian noise, similar to amplifier noise\revA{~\cite{mano2023modeling}}.
The amount of transceiver noise is obtained from the back-to-back BER-OSNR curve, which represents the relationship between input OSNR and pre-FEC BER in a measurement setup without fiber propagation.
Traditional BER-OSNR measurements require multiple optical devices, such as optical filters, amplifiers, spectrum analyzers, and ASE noise sources.
In contrast, by exploiting the additive nature of Gaussian noise and the internally generated noise induced by changes in input power, rather than externally injected ASE noise, the BER-OSNR curve can be estimated by using only a VOA\@.
This approach significantly reduces the measurement cost of transceiver characterization.

\subsection{Transponder virtualization related technologies}

Virtualizing transponders and delegating their control and management to other operators requires a Network Operating System (NOS) architecture that supports resource isolation and access control, as well as open control interfaces.

\subsubsection{Resource isolation and access control}

When delegating a subset of transponder line ports to other operators, it is necessary to isolate the associated resources and allow access only to that portion within the NOS\@.
Two approaches can be considered: OS-level virtualization using virtual machines and process-level separation using containers.
Computing nodes inside a data center, such as servers,  can employ both approaches because they have abundant computing resources.
However, since transponders are primarily designed for communication rather than computation and have limited computing resources, the lightweight container-based approach is advantageous.
For example, Telecom Infra Project (TIP) had been developing Goldstone, a container-based NOS~\cite{tip2025goldstone}.
We can achieve resource isolation and access control by implementing a container dedicated to controlling and managing specific line ports.

\subsubsection{Open Control Interfaces}

A virtualized transponder must be controllable and manageable by the delegated operator as intended.
If its control interface is vendor-specific and closed, the delegation would impose additional cost and delay on the operator.
Conversely, providing standardized open interfaces allows automation through APIs and seamless integration into existing ecosystems, thereby contributing to rapid service recovery.
In fact, both TIP and Open ROADM MSA have been discussing open interfaces for transponder control, where YANG is adopted as the data model and NETCONF as the transport protocol~\cite{openroadm}.
Such open interfaces are expected to facilitate the rapid establishment of optical paths by the delegated operator.

\section{Validation through field evaluations}
\label{sec:field-evaluations}

\begin{figure*}[ht!]
  \centering
  \includegraphics[width=0.7\linewidth]{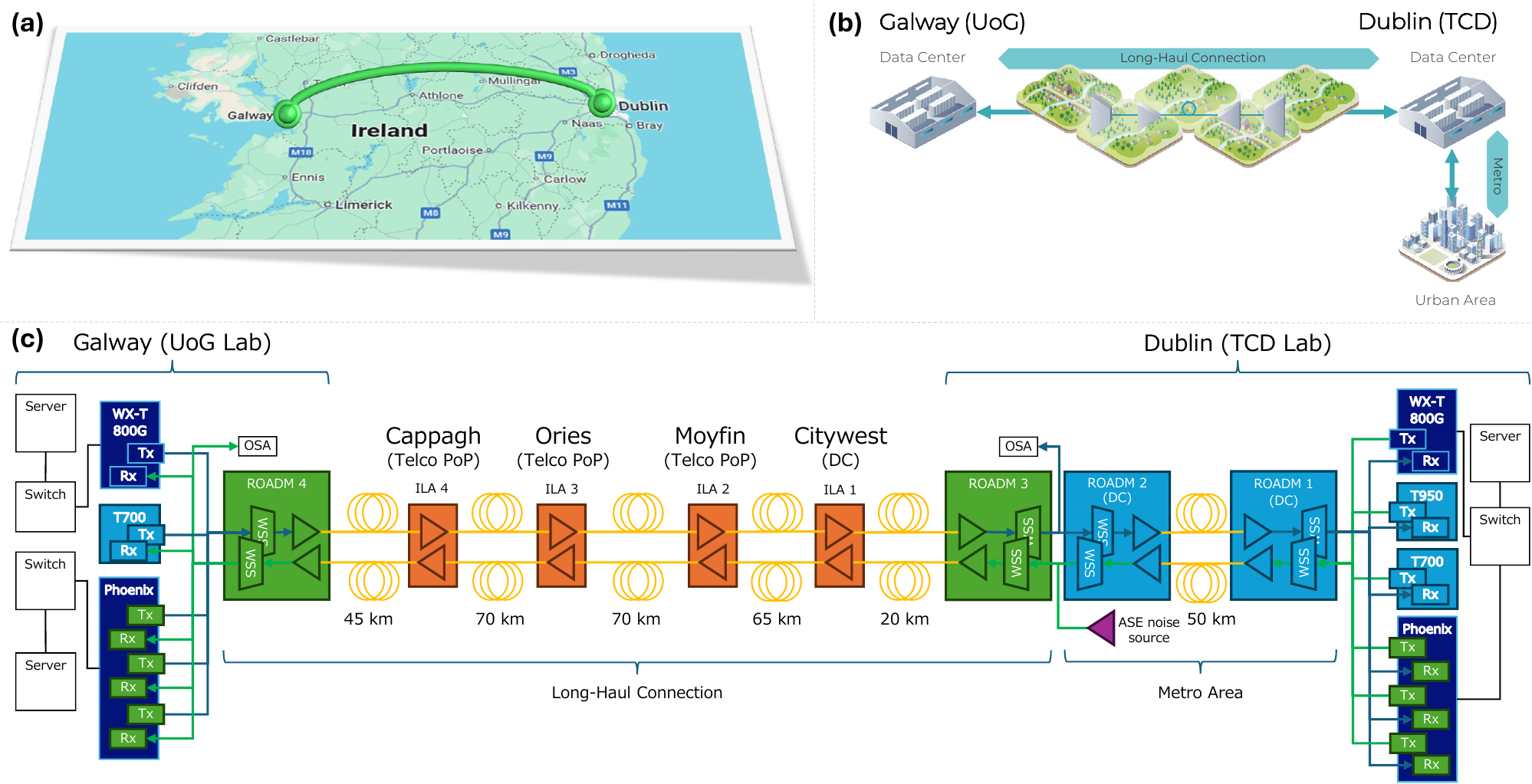}
  %\vspace{-4mm}
  \caption{
    Field-trial optical network connecting Dublin and Galway.
    (a) Geographical layout.
    (b) Use-case functional representation.
    The topology emulates a long-haul connection between two data centers and a short reach metro link.
    (c) Hardware setup.
    The multi-vendor, bidirectional network consisting of two line systems and three nodes.
    Dark fibers and amplification sites are deployed in the field alongside the production network.
    Color indicates vendor differences.
    %\vspace{-6mm}
  }
  \label{fig:use-case}
\end{figure*}

This section aims to experimentally prove the feasibility of agile resilience in disaggregated optical networks through a field trial using an OLS to connect sites roughly 300~km apart and the key enablers (Sec.~\ref{sec:key-enablers}).
The presented experiment assumes a post-disaster scenario (Sec.~\ref{sec:requirements-and-challenges}) in which the deployed equipment is borrowed from another operator.
Without dispatching personnel to each local amplification site, the physical characteristics of the fiber and amplifier were quantified and incorporated into a physical model.
%to virtually represent the transmission environment.
%
Using this calibrated model, multiple optical paths were deployed to carry traffic.
Subsequently, computing resources, specifically a database, were migrated using the newly established communication path.

\subsection{Setup}
Fig.\ref{fig:use-case} depicts the field-trial optical testbed deployed between the two cities of Dublin and Galway, designed to validate data center applications (Fig.\ref{fig:use-case}-a).
The layout reflects two representative connectivity scenarios: a long-haul link interconnecting data centers situated in different regions of the country, and a metro-scale segment serving the needs of an urban area (Fig.\ref{fig:use-case}-b).
The setup follows a multi-vendor strategy and realizes a bidirectional linear optical topology, incorporating two line systems and three intermediate nodes (Fig.\ref{fig:use-case}-c).
The evaluation was performed over dark fibers with amplification sites deployed in the field alongside the operational production network, thereby providing realistic conditions for assessing the proposed solution.

Each line system is defined by a specific ROADM-to-ROADM connection.
The 280~km long-haul link consists of five fiber spans including four in-line amplifiers (ILAs).
The two terminals of the long-haul link are equipped with OSAs and an ASE noise source to perform the OLS calibration procedure and fill spectrum for troubleshooting and in-service transmission operation.
The short-reach link is emulated in the laboratory using a single fiber spool of 50~km.
The portion of WDM spectrum implementing the data traffic is created by means of transceivers supporting 800G and 400G bit rates.
The 800G wavelengths are obtained from NEC SpectralWave WX-T 800G and Fujitsu 1FINITY T950, while the 400G ones use NEC SpectralWave Phoenix WX-T\@.
NEC NOS on SpectralWave has been used for transponder virtualization testing.
This commercial NOS, based on TIP Goldstone NOS and an open control interface, deploys a dedicated container for the management of each line port.
The rest of the full C-band spectrum is filled with 50-GHz dummy channels on a 100-GHz grid.

\subsection{Procedure and time record}
We successfully provisioned optical paths between two sites approximately 300~km apart and performed a database migration over the established optical link.
The experimental procedure was as follows:
\begin{enumerate}
  \item \textit{Transceiver Characterization (30 minutes)}\\
        The transceivers' performance characteristics were assessed at both endpoints.
  \item \textit{DLM for Fiber Parameter Extraction (1 hour total)}
        \begin{itemize}
          \item \textit{Measurement (20 minutes):} The received signals are captured at the receiver transponder.
          \item \textit{Analysis (40 minutes):} The collected data was analyzed to extract key fiber characteristics such as attenuation, dispersion, and nonlinearities.
        \end{itemize}
  \item \textit{OLS Calibration and Optimization (3.5 hours total)}
        \begin{itemize}
          \item \textit{Measurement (2.5 hours):} The OLS was measured under different operating conditions to estimate the transmission physical parameters of the link.
          \item \textit{Analysis (1 hour):} The system parameters were estimated and the amplifier configurations were optimized based on the calibration results to ensure stable and high-quality transmission.
        \end{itemize}
  \item \textit{DLM Visualization of Optimized Line System (1 hour)}\\
        A second DLM run was performed to visualize and validate the optimized line system configuration.
  \item \textit{Transponder Configuration Optimization (2 minutes)}\\
        Transponder settings were designed to define the corresponding optical path according to the line system QoT estimation.
  \item \textit{Database Migration (10 minutes)}\\
        A 25~GB database was initialized and migrated over the provisioned optical path.
\end{enumerate}

\begin{figure}[!b]
  %\vspace{-6mm}
  \centering
  \includegraphics[width=0.45\columnwidth]{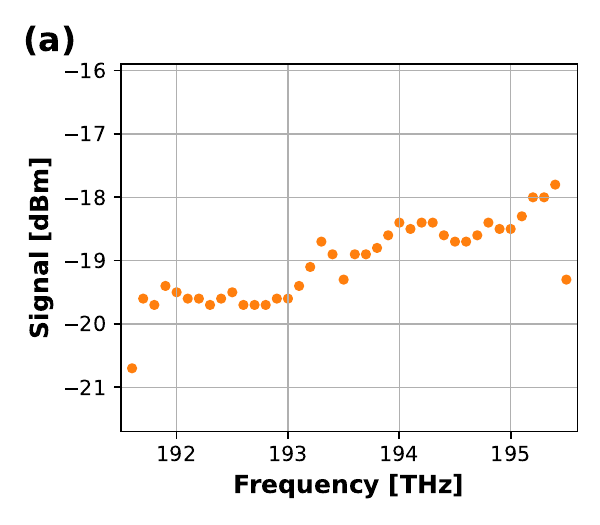}
  \includegraphics[width=0.45\columnwidth]{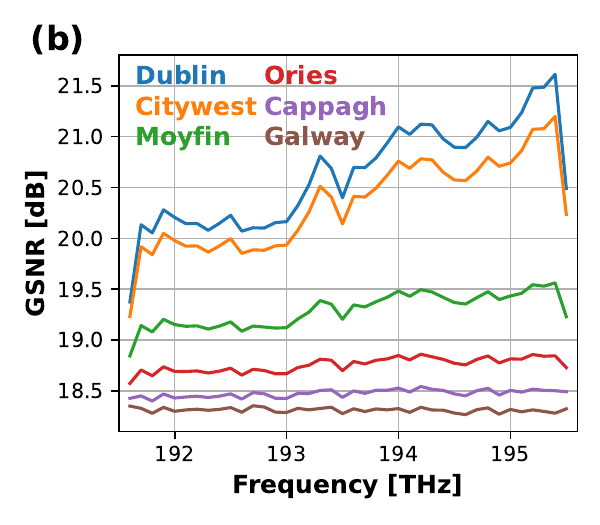}
  %\vspace{-2mm}
  \caption{
    Calibration results for the OLS from Dublin to Galway.
    (a) Computed optimal launch power before booster.
    (b) Computed accumulated GSNR after each EDFA\@.
    The optimal launch power is computed to have the highest flattened GSNR spectrum at the end of the line.
  }
  \label{fig:ols-calib-results}
\end{figure}

\begin{figure}[!t]
  \centering
  \begin{minipage}[b]{0.65\columnwidth}
    \centering
    \includegraphics[width=\textwidth]{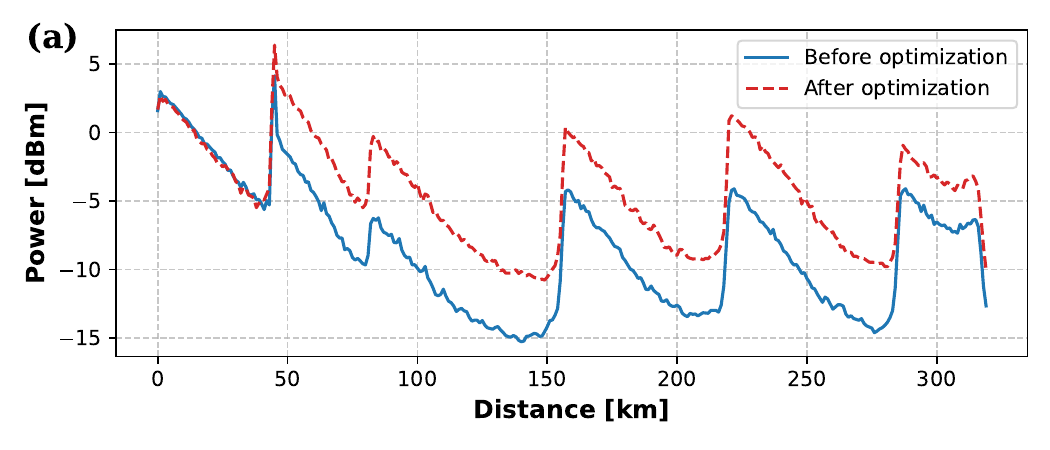}
  \end{minipage}
  \raisebox{18pt}{
    \begin{minipage}[b]{0.3\columnwidth}
      \centering
      \leftline{\textbf{\scriptsize (b)}}
      \includegraphics[width=\textwidth]{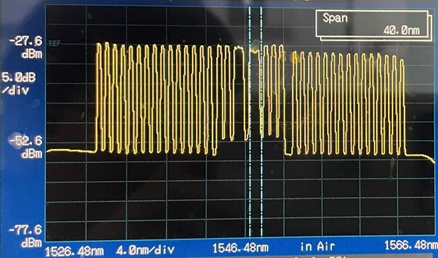}
    \end{minipage}
  }
  %\vspace{-6mm}
  \caption{
    DLM validation.
    (a) Analyzed power profile.
    The vertical axis and horizontal axis show absolute power and distance from the Galway site, respectively.
    The blue solid line and the red dotted line show before/after optimization.
    (b) Received WDM spectrum during validation, consisting of DLM 800G (130~GBd), 400G (63.1~GBd) and ASE noise loading (50~GHz) channels.
  }
  \label{fig:analyzed-pp_wdm-spectrum}
\end{figure}

% Launch Power Design
After performing the parameter calibration of the physical model~\cite{borraccini2025optical},
the model was used to design the optimal amplifier configurations, including gain and tilt settings, and the optimal WDM channel launch power before the booster.
The optimization target is the generalized signal-to-noise ratio (GSNR),
%~\cite{curri2022gnpy},  to save refs.
which allows us to relate the optical transmission performance of the system, or QoT, to transceiver BER~\cite{mano2024measuring}.
The design simulation results of the launch power optimization are reported in Fig.~\ref{fig:ols-calib-results}, illustrating the per-channel launch power configuration and the GSNR spectrum evolution after each span of the long-haul link.
The GSNR was flattened along the full C-band spectrum to ensure QoT uniformity for all optical paths crossing the link,
%~\cite{borraccini2025qot} to save refs.
assuming a scenario of meshed network or data center exchange~\cite{nishizawa2024fast}.
Fig.~\ref{fig:analyzed-pp_wdm-spectrum} (a) shows the power profile along the link from Galway to Dublin obtained with DLM, comparing the results before and after OLS calibration (black and red dotted lines, respectively).
% may be cut to save words since no supporting results
This tool enables visualization of the channel power evolution, allowing the detection of irregular losses within fiber spans and the localization of power profile variations along the spatial coordinate over time.

\subsection{Results and Discussions}
% GSNR Validation
The QoT of the long-haul link was validated after provisioning two 800G and four 400G channels and measuring the corresponding pre-FEC BER\@.
Figure~\ref{fig:analyzed-pp_wdm-spectrum} (b) reports the measured WDM spectrum after optical path deployment.
The difference between the estimated QoT at the design stage and the measured QoT after path establishment was conservative with a margin of roughly 1~dB.
%
%Descripton for key enabler B
%
% RESULTS
The results showed that the entire process, including system characterization, path design and establishment and database migration, was completed in approximately \textit{6 hours}.
The migration was completed without errors, confirming the stability and performance of the optical link.

\revB{%
  While the 6-hour provisioning time is faster than manual configuration, there is an inherent trade-off between provisioning speed and optimization depth.
  The optimal balance in disaster recovery depends on factors such as link conditions, operational urgency, and specific requirements from users or authorities.
  Simpler optimization methods could save several hours but may result in suboptimal configurations.
  In particular, they may not adequately cope with nonlinear optical effects or unexpected loss variations in newly constructed lines using unknown dark fibers, thereby posing risks of link establishment failure or severe instability.
  This paper prioritizes the reliability of service restoration, adopting an approach that combines full characterization and in-depth optimization realized by DLM and OLS calibration.
  Nevertheless, further reducing provisioning time for scenarios where speed is the primary concern remains an important topic for future work.

  Note that, as the temporary line system is newly constructed specifically for disaster recovery and carries no pre-existing traffic, the OLS calibration can be executed without causing any service disruption.
}%

\section{Future research}
\label{sec:future-researches}
Future work will focus on further enhancing resilience by leveraging free-space optical (FSO) communications via satellites, which are inherently less susceptible to natural disasters.
Another promising approach is the deployment of high-capacity terrestrial FSO terminals to establish temporary short links for emergency connection among data centers when fiber links are unavailable or congested.
Also, given the open-ended nature of disaster recovery, where tasks such as identifying feasible migration routes under severe constraints must be completed within a limited time, AI-driven solutions offer significant potential.
Specifically, specialized AI and large language model (LLM) agents can be employed to narrow the search space and automate the initial stages of emergency route leasing and provisioning, thereby reducing response time and improving overall agility.

\section{Conclusion}
\label{sec:conclusion}

Given that recent environmental shifts have amplified disaster unpredictability, the effectiveness of traditional redundancy-based approaches is failing.
To address this, we proposed agile resilience, which enables dynamic resource reallocation across domains and operators.
We used a field-deployed system to demonstrate characterization, path provisioning, and service migration within 6 hours, confirming our proposal's feasibility.
Future work includes AI-driven automation and satellite-based optics, positioning agile resilience as a critical foundation for a sustainable information society.

\section*{Acknowledgments}
This work is supported in part by a research grant from Taighde Éireann (Research Ireland) under 22/FFP-A/10598 (Twilights), 18/RI/5721 (Open Ireland), 13/RC/2077 P2 at CONNECT: the Research Ireland Centre for Future Networks, and from National Institute of Information and Communications Technology (NICT), Japan, under JPJ012368G50201.

\bibliographystyle{IEEEtran}
\bibliography{IEEEabrv,references}

\section*{Biographies}
% Authors’ bios are not to exceed 150 words each.

\begin{IEEEbiographynophoto}{Toru Mano}
  is a research engineer at NTT Network Innovation Laboratories.
  He received his B.E. and M.E. degrees from the University of Tokyo in 2009 and 2011, respectively.
  He joined NTT Laboratories in 2011.
  He received his Ph.D. degree in computer science and information technology from Hokkaido University in 2020.
  %His research interests include network architectures, network optimization, and network softwarization.
\end{IEEEbiographynophoto}

\begin{IEEEbiographynophoto}{Hideki Nishizawa}
  is a Senior Research Engineer at NTT Network Innovation Laboratories, Japan. He received B.E. and M.E. degrees in Physics from Chiba University in 1994 and 1996, respectively, and a Ph.D. in Information Science and Technology from Hokkaido University in 2025.
\end{IEEEbiographynophoto}

\begin{IEEEbiographynophoto}{Takeo~Sasai}
  received the B.E., M.E. and Ph.D. degrees from the University of Tokyo, Japan, in 2015, 2017, and 2025 respectively. In 2017, he joined NTT Laboratories, Yokosuka, Japan.
  %His research interests include physical layer monitoring, digital signal processing, and nonlinear modeling for optical fiber communication.
\end{IEEEbiographynophoto}

% copy from https://www.ntt-review.jp/archive/ntttechnical.php?contents=ntr202112fa10.pdf&mode=show_pdf
\begin{IEEEbiographynophoto}{Soichiroh Usui}
  is a senior research engineer NTT Network Innovation Center.
  He received a B.S. in industrial engineering and management from Tokyo Institute of Technology in 1999.
  He joined NTT the same year and was involved in corporate business and network services development.
\end{IEEEbiographynophoto}

\begin{IEEEbiographynophoto}{Dmitrii Briantcev}
  is a research fellow at the CONNECT Centre, Trinity College Dublin. He received the B.Sc.\ in Radiophysics from Saint Petersburg State University (2018) and the M.Sc. (2020) and Ph.D. (2023) in Electrical and Computer Engineering from KAUST, Saudi Arabia.
  % He works on optical communications and digital twins, integrating simulation, data-driven channel modeling, and testbed validation.
\end{IEEEbiographynophoto}

\begin{IEEEbiographynophoto}{Devika Dass}
  is a research fellow at the CONNECT Centre, Trinity College Dublin.
  %Her research interest encompasses exploring higher-order data modulation for converged optical metro-access networks, analog Radio-over-Fiber and Free-space optics.
  She earned her Ph.D. in 2023 from Dublin City University, master’s in Communication Engineering from VIT University, Vellore, India (2017) and undergraduate in Electronics and Telecommunication from Amity University, India (2014).
\end{IEEEbiographynophoto}

\begin{IEEEbiographynophoto}{Brandt Bashaw}
  is a Ph.D. student at the CONNECT Centre, Trinity College Dublin. He received his B.Sc.\ in Electrical Engineering from Brigham Young University in 2023. His research interests include classical optical communications, quantum optics, and quantum computing.
\end{IEEEbiographynophoto}

\begin{IEEEbiographynophoto}{Eoin Kenny}
  is the Innovation, Research and Development manager at HEAnet (Ireland's National Education and Research Network).
  %His research interests are in the areas of advanced network architectures, quantum communications, distribution of timing services, fibre sensing, and control and management planes for future networks.
  He received a degree in Electronic Engineering from University College Dublin in 1993.
\end{IEEEbiographynophoto}

\begin{IEEEbiographynophoto}{Marco Ruffini}
is full professor at Trinity College Dublin, leading the OpenIreland lab and the Optical and Wireless networks research group.
%His main research is in the area of Intelligent networks, where he carries out pioneering work on the convergence of fixed-mobile and access-metro networks, AI-based control and digital twins of optical networks, quantum networking and fibre sensing. He has been invited to share his vision through several keynote and talks at major international conferences across the world.
He authored over 220 international publications,
%10 patents, contributed to industry standards
and secured research funding for over € 14 milion, and contributed his novel virtual Dynamic Bandwidth Allocation (vDBA) concept to the BroadBand Forum standardisation body.
\end{IEEEbiographynophoto}

\begin{IEEEbiographynophoto}{Yoshiaki Sone}
  is a senior research engineer at NTT Network Innovation Laboratories. He received his M.E. degree in electronics engineering from Tohoku University, Miyagi, in 2003. In 2003, he joined NTT Laboratories.
  %to focus his research on network engineering technologies for optical transport networks.
\end{IEEEbiographynophoto}

\begin{IEEEbiographynophoto}{Koichi Takasugi}
  received the B.E. degree in computer science from Tokyo Institute of Technology in 1995, the M.E. degree from Japan Advanced Institute of Science and Technology in 1997 and Ph.D. in engineering from Waseda University in 2004.
  In 1997, he joined NTT Laboratories, Japan.
  %, where he has been engaged in research on wireless and optical transport networks.
\end{IEEEbiographynophoto}

% from https://ieeexplore.ieee.org/document/10747065
\begin{IEEEbiographynophoto}{Daniel Kilper}
is Professor of Future Communication Networks and SFI CONNECT Centre Director at Trinity College Dublin, Ireland.
%He holds an adjunct faculty appointment at Columbia University Data Science Institute and College of Optical Sciences, University of Arizona.
%He is CTO and co-founder of Palo Verde Networks, Inc., and is the Green Internet and Service Provisioning topical area editor for IEEE Transactions on Green Communications and Networking.
%He co-chairs the IEEE International Network Generations Roadmap (INGR) Optics Working Group.
He received M.S. (1992) and Ph.D. (1996) degrees in physics from the University of Michigan.
%From 2000 to 2013, he was a member of the technical staff at Bell Labs.
% bio must be less than 150 words
%His work has been recognized with a NIST Communication Technology Lab Innovator Award and a Bell Labs President’s Gold Medal Award, and he served on the Bell Labs Presidents Advisory Council on Research.
%He holds 13 patents and has coauthored 6 book chapters and more than 170 peer-reviewed publications.
His research is aimed at solving fundamental and real-world problems in communication networks, addressing interdisciplinary challenges for smart cities, sustainability, and digital equity.
\end{IEEEbiographynophoto}

\vfill

\end{document}